\documentclass[epsf,12pt]{iopart}

\usepackage{epsfig}
\usepackage{graphicx}

\begin{document}

\title{Dusty Plasma Liquids}

\author{Chia-Ling Chan, Ying-Ju Lai, Wei-Yen Woon, Hong-Yu Chu and Lin I}

\address{Department of Physics and Center for Complex Systems, National
 Central University, Chungli, Taiwan 32054, R.O.C.}

\ead{lini@phy.ncu.edu.tw}

\begin{abstract}
The dusty plasma liquid formed by micro-meter sized particles negatively charged and suspended in a low pressure discharge background is a good candidate to study the generic spatio-temporal dynamical behaviors at the kinetic level through direct optical video-microscopy, because of the sub-mm interparticle spacing and the slow time scale. In this paper, starting from the basic picture of the avalanche type stick-slip cooperative hopping process under the interplay of mutual coupling and the stochastic thermal agitation, we briefly review our recent study on particle micro-motion and the associated structural rearrangement. The effect of mesoscopic confinement and external slow shear drive are also discussed. \end{abstract}

Depending on the observation scale, the dynamics of liquids can be roughly approached by two different ways. Macroscopically, liquid is a continuous extended system with homogeneous structure and exhibits plastic deformation under external shear \cite{1}. However, down to the atomic scale, the discreteness and thermal fluctuations plays important roles. The continuous mean field approach with partial differential equations can no longer be used to describe the dynamical behavior of the discrete system. The system can be treated as a strongly coupled many body system under stochastic thermal agitation. Namely, the dynamics should be described by the coupled Langevin equations instead of the Navier-Stoke equations.  Due to the lack of tools for direct visualization at the small atomic scale, the general understanding of the micro-behaviors of liquids seems to be limited to the intuitive impression of disordered micro-structure and Brownian type motion induced by the background thermal noise. Actually, the liquid shares the similar inter-particle distance to the solid. Its strong mutual coupling and the interplay with thermal agitation and discreteness could lead to many interesting dynamical behaviors in cooperative particle micro-motion and the heterogeneous structure rearrangement, i.e. coherence out of disorder.  Without the proper tool for direct microscopic observation, it is difficult to experimentally address the following issues at the kinetic level: 1) The generic statistical behaviors of the above cooperative excitations. 2) The generic behavior of non-Gaussian dynamics and anomalous diffusion at the kinetic time scale. 3) Whether there exists some similar universal features to other strongly coupled complex system under stochastic or slow drives, e.g. self-organized criticality (SOC) behavior. \cite{2}. 4) The generic behavior of liquid confined under mesoscopic confinement. 5) The microscopic response to the external shear and the microscopic origin of viscosity. In this paper, we briefly review our recent experimental studies on these issues using a toy system, dusty plasma liquid, to mimic and construct a microscopic picture for the micro-behaviors in liquid through direct visualization under the proper scale (sub-mm inter-particle spacing and a few tens of thermal relaxation time).  

Dusty plasma can be formed by suspending micrometer sized fine particles in weakly ionized glow discharge \cite{3,4}. The strongly coupled mutual Coulomb coupling due to the strong negative charging of the dust particles (about ten thousand electrons per particle) can turn the system into crystal or liquid state with sub-mm interparticle spacing. The capabilities of directly visualizing particle micro-motion over a broad range of time scales using optical microscopy make it a good platform to explore many interesting dynamical behaviors \cite{5,6,7,8}. In our studies, a cylindrical symmetrical system as described elsewhere is used \cite{5,6}.  A weakly ionized discharge ($n_e$ $\sim$ $10^9$ $cm^{-3}$) is generated in 250 mTorr Ar gas using a 14 MHz rf power system.  A hollow cylindrical cell 30-mm in diameter is placed on the center of the bottom electrode to trap polystyrene particles at 7 $\mu$m diameter. It can also be replaced by rectangular cells with narrower width to study the lateral mesoscopic confinement effect. Particles are confined by the plasma sheath adjacent to the surrounding wall. Vertically, the suspended dust particles are aligned with eight particles for each chain by the wake field effect of the vertical ion flow. Particles in the same vertical chain move together horizontally.  It forms a quasi-2D system.  The particle positions in the horizontal mono-layer illuminated by a thin laser sheet are monitored through digital video optical microscopy.  The mean inter-particle distance $a$ is 0.3 mm.  

Fig. 1(a) to (d) show the typical particle trajectories at different time interval. Although the long time motion of the particle look quite disordered (Fig. 1(a)), interesting patterns are observed at 30 sec time interval. As shown in Fig. 1(b), particle motion can be roughly divided into the caged motion in the ordered triangular lattice type domains, and the string- and vortex-type hopping. In a cold 2D liquid melted from a triangular lattice, the inter-particle distance remains similar and most particles are still surrounded by six nearest neighbors (see the particle positions of particles in Fig. 1(d)) \cite{5}. However, unlike the solid in which particles only exhibit smaller amplitude oscillation in their own caging site, the accumulation of constructive perturbation from the stronger thermal agitation can make the particle overcome the caging barrier induced by the topological constraint from the nearest neighbors \cite{9,10}. Namely, particles exhibit alternate caged and hopping motion, i.e. stick-slip type motion. The Coulomb coupling provides a channel to propagate the information between particles. On one hand, it tends to generate ordered state with triangular structure. On the other hand, it can transfer the kinetic energy of hopping particles to the neighboring particles and induces cooperative hoppings in the form of strings or vortices involving a small number of particles. Fig. 1(e) illustrates the cooperative hopping in the xyt space. The vertical lines with small amplitude wiggling correspond to the caged state and the tilted lines correspond to the hopping state. Hopping ceases when the kinetic energy is quickly dissipated to the particles in the background network. Fig. 1(g) shows the histogram of particle hopping events. $\Delta t_F$ and $\Delta D_F$ are the persistent time the traveling distance for each hopping event respectively, and $a$ is the mean inter-particle distance.  Usually, a particle only travels about one $a$. The distorted local structure from the ordered triangular lattice can be rearranged and regain ordering after releasing the strain energy through hopping.  

\begin{figure}
\begin{center}
\epsfxsize=12cm
\epsfbox{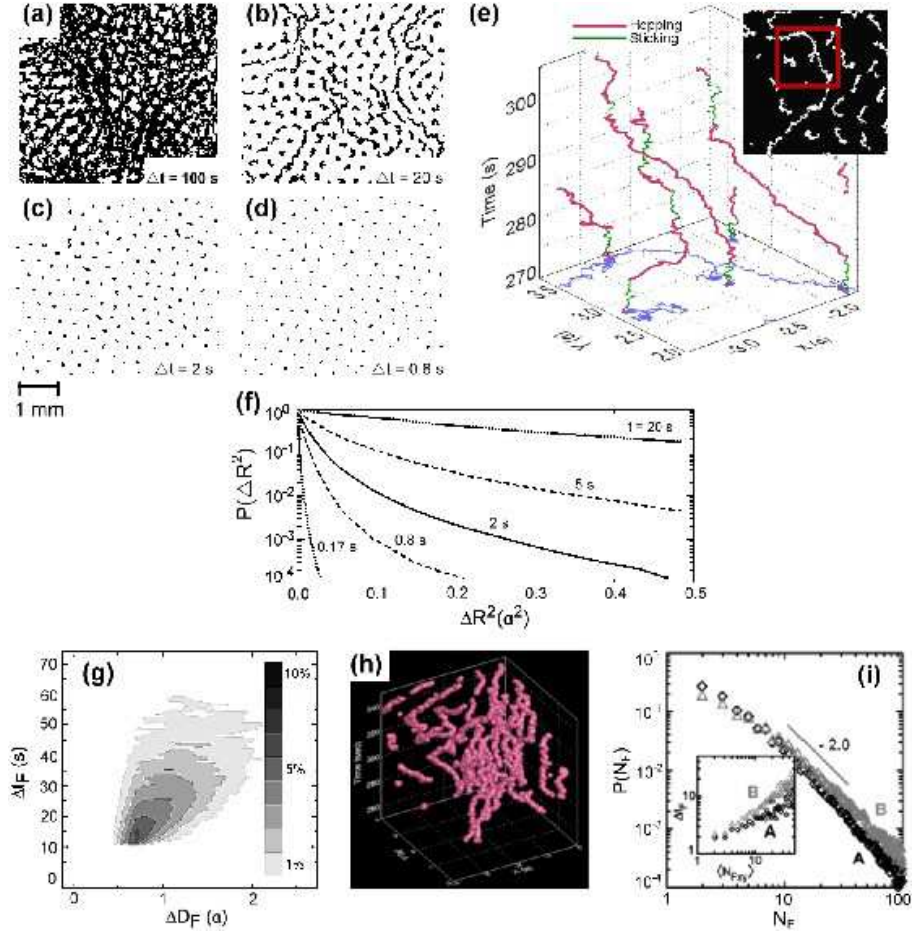} 
\end{center}
\caption{\label{fig1.eps} 
(a) to (d) The trajectories of particles in a cold liquid state with decreasing time intervals. (e) The 3D particle trajectories in xyt space corresponding to the upper right box of the 2D plot which is magnified from the upper left corner of (b). The red and green trajectories correspond to the hopping and caged states, respectively; and the blue lines are the corresponding projections on the xy plane. (f) The histograms of the particle mean square displacement $P(\Delta R^2)$. $a$ is the mean inter-particle distance. A good Gaussian distribution should follow a straight line. The upward banding of the histograms manifest the contribution from the nonequilibrium fast hopping particles. (g) The histogram of particle hopping events. $\Delta t_F$ and $\Delta D_F$ are the persistent time and the traveling distance for each hopping event respectively. (h) A typical 3D picture in xyt space showing the excitation of fast hopping particles (pink balls, the center of the balls is at the particle position and the diameter of the ball is 1 $a$) appear in the form of clusters. (i) The histograms of $N_F$, the size of the clusters of fast hopping particles in the xyt space normalized by the total number of clusters in the xyt space from run A and B. Run A is the cold liquid run generating the data shown in the figures above and run B is another hotter run.  The inset shows the power law relation between the temporal span $\Delta t_F$ and the averaged spatial span $\langle N_{Fxy} \rangle$ = $\langle N_F/\Delta t_F\rangle$ of the fast-particle clusters.  } 
\end{figure}

For a complicated system with such a large number of degrees of freedom, many information about motion and structure can be measured. For particle motion, we can easily measure the histogram of particle displacement. Intuitively, we might expect that particle exhibit Brownian type random motion with Gaussian type displacement histogram. Namely, the semi-log plot of P($\Delta R^2$) versus $\Delta R^2$) should follow a straight line. However, the curve bends upward in the intermediate time scale ($\tau$ around a few second)(Fig. 1(f)). It manifests the existence of the high-speed non-Gaussian tail. It is interesting to note that the mean square displacement (MSD = $\langle\Delta R^2\rangle$) versus time interval $\tau$ plot exhibits anomalous diffusion with the scaling exponent ($\alpha$) deviating from 1 (1 is for normal diffusion) \cite{6}. At short time interval (e.g. the trajectories in Fig. 1(d), particle motion is dominated by the small amplitude high frequency oscillation (see the fast wiggling in the trajectory plot in xyt space). It leads to the anti-persistent diffusion. For the intermediate time scale (about a few sec), sufficient constructive perturbation can be accumulated which leads to persistent hopping and $\alpha$ $>$ 1. In this regime, the fast hopping particles cause the upward bending of the curve in the displacement histogram plot. However, in the regime of a few tens of second, the successive random phase hoppings wash out the motion memory and lead to the normal diffusion with Gaussian type displacement histogram. Namely, both caged and hopped motions are consequences of the competition among discreteness, Coulomb coupling, and thermal kicks. Coherence can be partially preserved in small spatio-temporal scale. Figure 1(f) further locates the sites of hopping events in xyt space. Avalanched type excitation of hoppings in the form of clusters involving different number of particles can be observed. Unlike the intuitively expected uncorrelated distribution, the cluster size follows a scale-free power law distribution with slope around 2.0. (Fig. 1(i)). Note, the similar scale-free power law distribution have been observed for many strongly coupled sub-excitable systems under external random drives \cite{11,12}. The cold 2D dusty plasma liquid is a paradigm for the SOC behavior at the microscopic level. 

With the basic building blocks of slick-slip type caged and cooperative hopping under the competition between mutual coupling and thermal agitation, we can easily expect that increasing temperature will speed up hopping, i.e. shortening the transition time scale from the anti-persistent to persistent diffusion. It is interesting to note that similar exponents for the persistent diffusion and the scaling law of the clusters size distribution of fast hopping particles to the colder liquid runs are preserved (Fig. 1(i))\cite{6}.
 
\begin{figure}
\begin{center}
\epsfxsize=14cm
\epsfbox{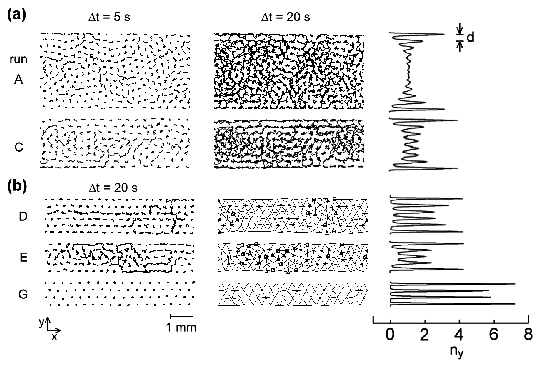} 
\end{center}
\caption{\label{fig2.eps} (a) The 5 and 20 sec particle trajectories of the cases at $W$ = 11 and 7 $d$. (b) The 20 sec particle trajectories and the typical triangulated structures with defects (triangles and squares represent the 5- and 7-fold defects respectively and particle trajectories (20 sec) at small $W$ for runs with narrower gap width $W$ = 5, 4.5 and 3 $d$. The right columns for (a) and (b) show the corresponding transverse density distribution $n_y$. The diffusion is anisotropic in the three outmost layers. Reducing the gap width causes the transition to the layered structure through the entire gap and slowing down of the dynamics.  The improper gap width deteriorates the good packing even when the gap width is reduced (e.g. at $W$ = 4.5 $d$).  } 
\end{figure}

We can further ask how the motion behaves when the system size goes down to the mesoscopic scale, induced by the effect of finite boundary under confinement, which is a hot topic for many fields such as nano-science, tribology, biology, etc.. \cite{13,14,15}.  Figure 2(a) and (b) show the typical particle trajectories, the corresponding transverse particle density distributions $n_y$ versus the transverse distance $y$, for the liquid confined in narrow gaps with decreasing width \cite{17}.  The oscillating $n_y$ manifest that particles form layered structure around both boundaries. The decaying amplitude indicates that layering is gradually deteriorated and the positions are then randomized while moving away from the boundary. The layered structure can extend through the entire gap, associated with the slow particle dynamics, while the gap width is reduced below 7 interlayer distance. Namely, the presence of finite boundaries sets up topological constraints and reduces entropy. It lines up particles and suppresses particle transverse hopping.  This effect last about 3 interparticle distance, which is about the same as the correlation length for the pair correlation function. The diffusion is therefore anisotropic in the boundary regions about 3 interparticle distance in width. In the center region, the hopping is still isotropic, similarly to the bulk liquid. While reducing the gap width down to 7 interlayer distance, the layered structures from both boundaries pinches. Namely, the topological constraints from the boundaries and discreteness effect of particles suppress the pathway for particle structure rearrangement. It therefore slows down the microdynamics with much smaller transverse and longitudinal diffusion rate and the make the structure with better structural ordering.  This is also the microscopic origin of the observed sluggish behavior for the liquid in the mesoscopic gap. Our recent study demonstrates the formation of circular shells under the strong suppression of the inter-shell hopping in the few outmost shells around the boundary, and the isotropic hopping vortices in the center part \cite{16}. 

\begin{figure}
\begin{center}
\epsfxsize=12cm
\epsfbox{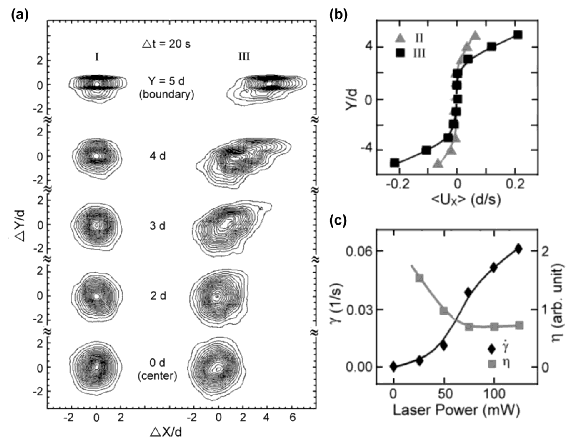} 
\end{center}
\caption{\label{fig3.eps} (a) The contour plots of the 2D histogram of particle displacement at 20 sec exposure time, for particles sitting in the longitudinal strips 1$d$ in width and at different transverse positions at 0 and 100 W laser power. (b) The transverse profiles of particle mean longitudinal velocity $V_x$. High shear rate outer bands with enhanced diffusion rates sandwiching a less perturbed center zone is observed. (c) The shear thinning with decreasing shear viscosity and enhanced diffusion rate as the shear stress increases. } 
\end{figure}

Now, we like to address the effect of external shear stress on the liquid motion by the radiation pressure from two parallel but counter-propagating laser beams along the opposite boundaries of a thin liquid about 11 $d$ in width \cite{18}. Figure 3(a) shows the contour plots of the 2D histogram of particle displacement for particles sitting in the longitudinal strips 1$d$ (i.e. the interlayer distance shown in Fig. 2(a) in width and at different transverse positions at 0 and 100 W laser power. The effect of boundary confinement is again manifested by the anisotropic contours for the outmost few layers. Applying shear not only causes the shift of the center position of plots but also enlarged the transverse as well as the longitudinal velocity fluctuation (i.e. the diffusion rate) for the few outmost layers. Unlike the linear velocity profile with uniform flow rate for the bulk liquid, shear banding with two high shear rate outer bands sandwiching a center low shear rate band is further evidenced by the mean longitudinal velocity profile in Fig. 3(b). The mean shear rate has an S shape response to the external stress, which manifests the shear thinning, i.e. decreasing viscosity with increasing shear stress (Fig. 3(c)). Since hopping is a nonlinear threshold type local process, the low hopping probability at the low stress limit is the physical origin of the high viscosity and lower diffusion rate in this regime.  Increasing the external stress further tilts the caging well and enhances the thermal assisted hopping. Through the disorder network, the external stress and the local stress from thermal agitation and hopping motion can also propagate transversely, which enhances the transverse diffusion in addition to the forward motion. Note that the width of the shear band is about the same as the layering width nearby the boundary, which is also the same as the typical hopping vortex size. Namely, the stress induced cooperative hopping induces avalanche type structure rearrangement associated with the local stress release involving a small number of particles shield the external stress and leaves a less perturbed center zone. Also note that in another experiment with a single laser beam pushing through the center of a very large 2D liquid, similar shear thinning with the similar shear band width was also observed \cite{5}.

The above results from particle trajectories, particle average velocity and particle diffusion reflect the information of particle position evolution with respect to the lab-frame. Moreover, through tracking particle trajectories, we can measure the local structural order and its evolution through measuring the local bond-orientation order (BOO) and defect excitation. BOO is defined as $\psi_6(r)$ = $\frac{1}{N_i}\sum_iexp(i6\theta_i)$, where $\theta_i$ is the angle of the vector from the particle at $r$ to its $i$th nearest neighbor, and $N_i$ is the number of the nearest neighbors  $\vert\psi_6\vert$ = 1 at the perfect 6-fold lattice sites and $<$ 0.4 for the defect sites, which are defined as the sites with nearest neighbor number deviating from six, respectively \cite{19}. As seen in figure 4 (a) and (b), The ordered lattice type structure is strongly distorted around the defect sites where the strain energy increases due to the accumulation of the stress from background thermal noise and external drive. Namely, the stick-slip hopping process can cause the structural rearrangement through defect excitation and relaxation. The structural rearrangement process exhibits avalanche type behaviors in the form of clusters, and they are closely correlated to the fast particle hopping string, as shown in figure 4(c). The cluster size distribution also follows a power law distribution \cite{18}.  The structural rearrangement can relax the local stress. It can shield the external shear drive and leaves a center zone with low perturbation, which is the microscopic origin of the shear banding observed above. 

\begin{figure}
\begin{center}
\epsfxsize=14cm
\epsfbox{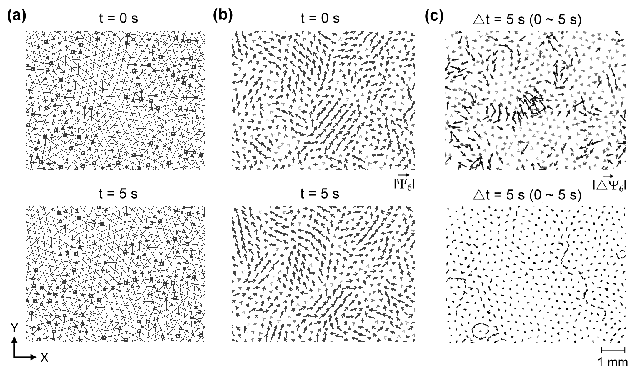} 
\end{center}
\caption{\label{fig4.eps} (a) and (b) 
The typical triangulated plots showing the change of defects with 5 sec seperation and the corresponding bond orientational order at different times. The length and angle of the vector in (b) correspond to $\vert \Psi_6\vert$ and $\theta_6$ respectively.  (c) The configuration for the change of $\Psi_6$ and the corresponding particle trajectories in 5 sec interval showing the structural rearrangement also occurs in the form of avalanche type clusters.  The excitations of defect (topologically disorder) alternately occur in different areas through the hopping process.  } 
\end{figure}

Finally, we like to show an interesting phenomenon of the formation of traveling plasma bubble through other nonlinear processes in the dusty plasma liquid \cite{20}. Figure 5 shows the side view of the quasi-2D dusty plasma liquid. A pulsed laser beam is focused at one of the suspended dust particle. The ablation on the particle generates an intense plasma plume which pushes the surrounding dust particles radially outwards. It leads to the formation of a bubble (void). Unlike the intuitive thought, the bubble does not collapse right away. It maintains its shape, propagates downward and exits the bottom sheath boundary. The particle trajectories show that the leading and trailing edge of the bubble are expelling and attracting regions respectively for particles. The bubble is actually permeable. Particles inside the bubble travel with much higher velocity than the surrounding particles. Our discharge is constantly driven by the external steady state rf power which provides a steady source for electron and ion generation. The electrons and ions are then recombined on the solid surfaces of the surrounding boundaries and the suspending dust particles. The expelling of dust particle by the initial plume reduces the electron depletion and generates a region with higher electron density, which leads to the higher plasma density through the enhanced ionization process. The outward ion flow is thereby enhanced and keeps the bubble from collapsing. The downward ion flow (toward the bottom electrode) of the background plasma breaks the symmetry and blows the bubble downward. Note the bubble travels about the same speed as the dust density wave.

\begin{figure}
\begin{center}
\epsfxsize=8cm
\epsfbox{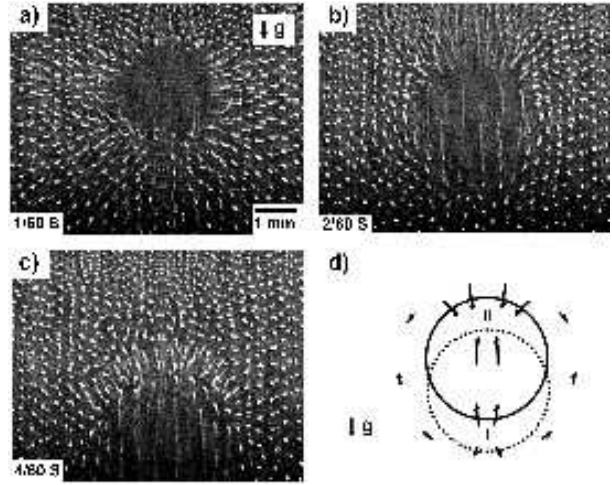} 
\end{center}
\caption{\label{fig5.eps} (a) to (c) 
The typical sequential snap shots of the side view of the downward traveling bubble in a dusty plasma liquid at 185 mTorr. The number in each lower left corner indicates the picture taken time (the laser pulse is fired at time zero) (d) The sketch showing the direction of particle velocities surrounding and inside the bubble while the bubble travels downward. The bubble represented by the initial solid circle advances downwards to the latter dotted circle by repelling particles from zone I and attracting particles into zone II with some phase delay. The length of each arrow is proportional to its speed. } 
\end{figure}

In conclusion, we have demonstrated that the dusty plasma is an interesting toy system for mimicking and understanding the micro-motion and structure at the kinetic level. On the other hand, it also exhibits interesting nonlinear phenomena through its own nonlinear process. At the kinetic level, the detailed interplay of the forces from the mutual coupling and the stresses from thermal agitation and other external slow drives is the key determining the fate of local motion and structure.  The system belongs to the category of the {\it integrate and fire} system. The integration of sufficient local stress from constructive perturbation induces hopping (firing). In addition to generating order, the mutual coupling also propagates the information of local excitation to the adjacent sites make the cooperative or avalanche type excitations, and quickly dissipate the excitation. The presence of straight boundary put topological constraint. It reduces the number of accessible states and kinetic pathways, which makes the stick-slip hopping anisotropic and causes the formation of layering transition with slow dynamics while the gap width is reduced to a few interparticle distance. Applying external stress tilts the caging well and further enhances the diffusion and reduces the viscosity. The relaxation of the stress through avalanche type structural arrangement induces the formation of the shear band.     

This work is supported by the National Science Council of the Republic of
China, under contract number NSC 92-0212-MOO8-042.

--------------------------------------------------------------------

\end{document}